\documentclass[3p,times,twocolumn]{elsarticle}

\usepackage{ecrc}
\usepackage{fancyhdr}
\usepackage{amsmath}
\usepackage{graphicx}
\usepackage{dcolumn}
\usepackage{bm}
\usepackage{amssymb}
\usepackage{epsfig}    
\usepackage{color}
\usepackage{slashed}

\volume{00}

\firstpage{1}

\journalname{Nuclear Physics B Proceedings Supplement}

\runauth{}

\jid{nuphbp}

\jnltitlelogo{Nuclear Physics B Proceedings Supplement}

\usepackage{amssymb}
\usepackage[figuresright]{rotating}

\begin{document}

\begin{frontmatter}



\dochead{}

\title{Radiative corrections to Higgs coupling constants in two Higgs doublet models}
%\footnote{This proceedings is based the research on Ref.~\cite{KKY yukawa}}


\author{Mariko Kikuchi \corref{r1}}

\address{Department of Physics, University of Toyama, 3190 Gofuku, Toyama 930-8555, JAPAN}

\cortext[r1]{This talk is based on the collaboration with Shinya Kanemura and Kei Yagyu~\cite{KKY yukawa}}

\begin{abstract}
A pattern of deviations in the Standard Model (SM) like Higgs boson ($h$) couplings
from their SM predictions depends on the structure of the Higgs sector and the Yukawa interaction.
In particular, in Two Higgs Doublet Models (THDMs) with a softly-broken $Z_2$ symmetry, 
different characteristic patterns of deviation in Yukawa coupling constants ($hf\bar{f}$) 
can be allowed depending on four types of Yukawa interactions. 
We calculate $hf\bar{f}$ coupling constants at the one-loop
level in all the types of THDMs.  
Even if there is no deviation in the $hf\bar{f}$ couplings at the tree level, 
they can deviate from the SM predictions by a few percent due to extra Higgs boson loop contributions. 
We find that if the deviations in the gauge couplings $hVV$ ($V=Z,~W$) are found 
with an enough large to be measured at the International Linear Collider (ILC), 
the scale factors for the $hf\bar{f}$ couplings do not overlap 
among the THDMs with four types of Yukawa interactions 
even taking into account the radiative corrections. 
Therefore, in such a case,
we can indirectly determine the type of the THDMs at the ILC even without information from direct searches of the
additional Higgs bosons.
%
%% A pattern of deviations in coupling constants of Standard Model (SM) like Higgs boson from their SM predictions depends on the structure the Higgs sector and the Yukawa interaction. 
%% In particular, in four types of Two Higgs Doublet Models (THDMs) with a softly-broken $Z_2$ symmetry, Yukawa coupling constants can realize to deviate in different characteristic patterns. 
%% %There are possibilities to determine the shape of the Higgs sector whether it is one of THDMs or others models by comparing theoretical predictions of coupling deviations with future precision data of coupling measurements at future collider experiments. 
%% We calculate Yukawa coupling constants of the SM like Higgs boson at the one loop level in all the types of the THDMs, and evaluate differences from those of the SM. 
%% Even these is no deviation due to the tree level field mixing, Yukawa couplings can deviate from the SM predictions by several percent due to one loop corrections of the extra Higgs bosons. 
%% We find that the scale factors for the Yukawa interactions do not overlap in different types of THDMs even in the case with maximum radiative corrections, if gauge couplings are different from the SM predictions large enough to be measured at the International Linear Collider (ILC).  
%% Therefore, in such a case, we can indirectly determine the type of the THDMs at the ILC even without information from direct searches of the additional Higgs bosons.
\end{abstract}

\begin{keyword}
 Extended Higgs sectors, Radiative corrections
\end{keyword}

\end{frontmatter}

% \linenumbers

\section{Introduction}
Although the standard model (SM) like Higgs boson ($h$) was discovered at the LHC experiment~\cite{ATLAS_Higgs, CMS_Higgs}, a lot of things are still unknown in the Higgs sector: e.g., what is the origin of negative mass term in the SM Higgs potential, whether the Higgs field is an elementary field or a composite field. 
Furthermore, we have not yet understood the shape of the Higgs sector. 
The minimal Higgs sector of the SM is just an assumption. 
There is no principle that only one isospin doublet field must be present. 
There are possibilities that the Higgs sector is extended, and all extended Higgs sectors have not excluded at all by the data of the LHC. 
On the other hand, we can say that the structure of the Higgs sector is strongly related to a scenario of the new physics beyond the SM, because a lot of models based on those scenarios introduce extended the Higgs sectors. 
Determining the structure of the Higgs sector by bottom up approach is one of the most effective procedure to establish the new physics.

In this talk, we consider a possibility to reconstruct the shape of the Higgs sector by coupling measurements of the SM like Higgs boson at future collider experiments. 
In general, in extended Higgs models, coupling constants of the SM like Higgs boson $h$ deviate from the predictions in the SM due to two kinds of effects. 
One is the effect of field mixing.
%, namely it is a deviation at the tree level. 
The other is the loop effect due to the extra Higgs bosons. A pattern of deviations in Higgs couplings depend on the number of the Higgs field, their representations and the mass of Higgs bosons in the loop. 
It is possible to discriminate extended Higgs sectors by using future precision data and comprehensively evaluating all coupling constants of $h$ in each model.

Within the relatively large uncertainties in the current LHC data ($\sqrt{s}=7, 8$ TeV, the integrated luminosity ($L$) is about 25 fb$^{-1}$), measured Higgs couplings seem to be consistent with the SM at both ATLAS~\cite{ATLAS_Higgs} and CMS~\cite{CMS_Higgs}. 
At the high luminosity LHC with $\sqrt{s}=14$ TeV and $L=3000$ fb$^{-1}$ (HL-LHC), deviations in the $h$ couplings from the SM predictions can be measured with expected accuracies about 5$\%, 10\%$ and $5\%$ for $hWW(hZZ)$, $hbb$ and $h\tau\tau$, respectively~\cite{Peskin,ILC white, TDR, HWGR,CMS note}. 
Moreover, at the future linear collider such as the ILC with $\sqrt{s}=500$ GeV and $L=500$ fb$^{-1}$ (ILC500), those can be tested by $1.1\%, 1.6\%$ and $2.3\%$, respectively~\cite{Peskin, ILC white, TDR, HWGR, Euro}. 
It is natural that we calculate theoretical predictions of these Higgs couplings with higher order contributions to compare with such precise coupling measurements at the ILC.

We here consider Two Higgs Doublet Models (THDMs) with a softly-broken $Z_2$ symmetry, which are often motivated in new physics models beyond the SM~\cite{THDM_rev}. 
Although flavour changing neutral currents (FCNCs) appear at the tree level in models with the multi doublet Higgs sector, 
we here avoid FCNCs at the tree level by imposing the $Z_2$ symmetry~\cite{Z2_sym} to the model.
Consequently, there are four types of models whose Yukawa interactions are different with each other~\cite{THDMs, Akeroyd}.
We call them Type-I, Type-II, Type-X and Type-Y THDMs~\cite{typeX, su_logan}. 
There are many new physics models with the THDM structure. 
For example, the neutrinophilic model~\cite{neutrinophilic} approximately has the Type-I THDM structure. 
In the Minimal Supersymmetric Standard Models (MSSMs), the structure of the Type-II THDM is required. There are radiative seesaw models ~\cite{radiative seesaw, Kajiyama} whose Yukawa interactions are corresponded to those of Type-X.

We calculate all Yukawa couplings with $h$ in four types of THDMs including electroweak radiative corrections at the one-loop level, and evaluate renormalized scale factors. 
In the Ref.~\cite{Hollik1} and Ref.~\cite{Hollik2}, the self coupling constant $hhh$ and Yukawa coupling constants have been calculated at the one-loop level in the MSSM, respectively. 
In THDMs with the softly-broken $Z_2$ symmetry, 
one-loop corrections to the gauge couplings $hVV$ ($V = Z, W$) and the $hhh$ coupling have also been studied in Refs.~\cite{KOSC}. 
However, all the $hf\bar{f}$ couplings have not been comprehensively analyzed including radiative corrections in the four types of THDMs. 
We also discuss how to discriminate the types of THDMs by combining theoretical predictions of Higgs couplings and precision measurements at the ILC.\footnote{
It is also important to study the direct searches for extra Higgs bosons at future collider experiments.
Possibilities of direct searches of THDMs are investigated in Refs.~\cite{direct, fingerp}}

%%%%%%%%%%%%%%%%%%%%%%%%%%%%%%%%%
\section{Model}
In the THDMs, there is an additional isospin doublet Higgs field $\Phi_1$ other than the one $\Phi_2$ in the SM. $\Phi_1$ and $\Phi_2$ receive non-zero vacuum expectation values (VEVs) $v_1$ and $v_2$, respectively. 
They satisfy the relation $v^2\equiv v_1^2+v_2^2 = (\sqrt{2}G_F)^{-1}$.  Five physical mass eigenstates (i.e., charged Higgs bosons $H^\pm$, a CP-odd Higgs boson $A$ and two CP-even Higgs bosons $h, H$ ) and unphysical three Numb-Goldstone bosons $G^\pm, G^0$ appear.

In general, FCNCs can appear at the tree level in models including multi Higgs doublet fields, because the Yukawa interaction matrix and the mass matrix of fermions cannot be simultaneously diagonalized. 
We should avoid FCNCs at the tree level due to constraints from flavour experiments. 
We here assume the model with a softly-broken $Z_2$ symmetry, 
so that each fermion can couple to only one of the Higgs fields.
If we assign the charge of the $Z_2$ symmetry to $\Phi_1$, $\Phi_2$, left-handed quark doublet, left-handed lepton doublet and right-handed up-type quark singlet fields as $+, -, + ,+$ and $-$, respectively, four types of Yukawa interaction appear depending on the way of the assignment of the $Z_2$ charge for right-handed fermions as shown in Table I.
We adopt to call the four types as Type-I, Type-II, Type-X and Type-Y ~\cite{typeX, su_logan}.

We consider the CP invariant case in this proceedings. Then the Higgs potential is given as 
 \begin{eqnarray}
 V&=m_1^2\Phi_1^{\dagger}\Phi_1 +m_2^2\Phi_2^{\dagger}\Phi_2 
   -m_3^2(\Phi_1^{\dagger}\Phi_2+\Phi_2^{\dagger}\Phi_1)\\\nonumber
   &+\frac{\lambda_1}{2}(\Phi_1^{\dagger}\Phi_1)^2
   +\frac{\lambda_2}{2}(\Phi_2^{\dagger}\Phi_2)^2
   +\lambda_3(\Phi_1^{\dagger}\Phi_1)(\Phi_2^{\dagger}\Phi_2)\\\nonumber
   &+\lambda_4(\Phi_1^{\dagger}\Phi_2)(\Phi_2^{\dagger}\Phi_1)
   +\frac{\lambda_5}{2}\left[(\Phi_1^{\dagger}\Phi_2)^2+(\Phi_2^{\dagger}\Phi_1)^2\right],
 \label{Higgs pote}
 \end{eqnarray}
where all parameters (namely $m_1^2$ - $m_3^2$, $\lambda_1$ - $\lambda_5$) are real parameters. 
(If we do not assume the theory to be CP invariant, $m_3^2$ and $\lambda_5$ are generally complex \cite{Higgs Hunter's}.) 
$m_3^2$ is a parameter which indicates the softly breaking scale of the $Z_2$ symmetry. 
These eight parameters in the Higgs potential are rewritten by the physical parameters; namely, masses of $H^{\pm}, A, H$ and $h$, two mixing angles $\alpha$ and $\beta$ which correspond to those among CP-even Higgs fields and charged (and CP-odd) Higgs fields, respectively, the VEV $v$ and the remaining parameter $m_3^2$. 
We here take $M^2$ instead of $m_3^2$; i.e., $M^2=\frac{m_3^2}{\sin\beta\cos\beta}$~\cite{KOSC}.

We mention scale factors of SM like Higgs boson couplings defined as $\kappa_X = \frac{g_{hXX}^\textrm{THDM}}{g_{hXX}^\textrm{SM}}$,
where $X$ is any field interacting with $h$. 
At the tree level, scale factors of gauge couplings ($\kappa_V$ ($V=Z,W$)) correspond to $\sin(\beta-\alpha)$. 
We define the SM like limit so that $\sin(\beta-\alpha)$ approaches to unity in the limit~\cite{THDM_decouple}. 
Scale factors of Yukawa couplings are summarized in TABLE I. 
Of course, in the SM like limit, all scale factors of Yukawa couplings become unity. 

\begin{table}
\begin{center}
{\renewcommand\arraystretch{1.2}
\begin{tabular}{|c|ccccccc|}\hline\hline
&
\multicolumn{7}{c|}{$Z_2$ charge}
%&\multicolumn{9}{c}{Mixing factor}
\\\hline
&$\Phi_1$&$\Phi_2$&$Q_L$&$L_L$&
$u_R$&$d_R$&$e_R$
%&$\xi_h^u$ &$\xi_h^d$&$\xi_h^e$&$\xi_H^u$&$\xi_H^d$&$\xi_H^e$ &$\xi_A^u$&$\xi_A^d$&$\xi_A^e$
\\\hline
Type-I &$+$&
$-$&$+$&$+$&
$-$&$-$&$-$
%&$\frac{\cos\alpha}{\sin\beta}$&$\frac{\cos\alpha}{\sin\beta}$&$\frac{\cos\alpha}{\sin\beta}$&$\frac{\sin\alpha}{\sin\beta}$&$\frac{\sin\alpha}{\sin\beta}$&$\frac{\sin\alpha}{\sin\beta}$&$\cot\beta$&$-\cot\beta$&$-\cot\beta$
\\\hline
Type-II&$+$&
$-$&$+$&$+$&
$-$
&$+$&$+$
%&$\frac{\cos\alpha}{\sin\beta}$&$-\frac{\sin\alpha}{\cos\beta}$&$-\frac{\sin\alpha}{\cos\beta}$&$\frac{\sin\alpha}{\sin\beta}$&$\frac{\cos\alpha}{\cos\beta}$&$\frac{\cos\alpha}{\cos\beta}$&$\cot\beta$&$\tan\beta$&$\tan\beta$
\\\hline
Type-X &$+$&
$-$&$+$&$+$&
$-$
&$-$&$+$
%&$\frac{\cos\alpha}{\sin\beta}$&$\frac{\cos\alpha}{\sin\beta}$&$-\frac{\sin\alpha}{\cos\beta}$&$\frac{\sin\alpha}{\sin\beta}$&$\frac{\sin\alpha}{\sin\beta}$&$\frac{\cos\alpha}{\cos\beta}$&$\cot\beta$&$-\cot\beta$&$\tan\beta$
\\\hline
Type-Y &$+$&
$-$&$+$&$+$&
$-$
&$+$&$-$
%&$\frac{\cos\alpha}{\sin\beta}$&$-\frac{\sin\alpha}{\cos\beta}$&$\frac{\cos\alpha}{\sin\beta}$&$\frac{\sin\alpha}{\sin\beta}$&$\frac{\cos\alpha}{\cos\beta}$&$\frac{\sin\alpha}{\sin\beta}$&$\cot\beta$&$\tan\beta$&$-\cot\beta$
\\\hline\hline
\end{tabular}}
\caption{Charge assignment of the softly-broken $Z_2$ symmetry  given in Ref.~\cite{typeX}.}
\label{yukawa_tab}
\end{center}
\end{table}

\section{Analysis}
In this section, we show the results of our numerical calculations. We calculate several Higgs coupling constants at one-loop level by the on shell renormalization scheme. 
The details of the renormalization are shown in Ref.\cite{KKY yukawa}. 
Using the Higgs couplings calculated with one-loop corrections, we evaluate the renormalized scale factors. 
In particular, we here discuss the renormalized scale factors of Yukawa couplings defined by
 \begin{eqnarray}
 \hat{\kappa}_f \equiv \frac{\hat{\Gamma}_{hf\bar{f}}[p_1^2,p_2^2,q^2]_\textrm{THDM}}
                            {\hat{\Gamma}_{hf\bar{f}}[p_1^2,p_2^2,q^2]_\textrm{SM}},
 \end{eqnarray}
where $\hat{\Gamma}_{hf\bar{f}}[p_1^2,p_2^2,q^2]_{\textrm{SM}(\textrm{THDM})},$ are the renormalized coupling constants in the SM (THDMs). 
We also consider constraints on parameter regions from perturbative unitarity and vacuum stability. 
Perturbative unitarity and vacuum stability are studied in Refs.~\cite{pv_THDM} and Refs.~\cite{vs_THDM}, respectively. 
We here take the external momenta to be masses of external particles; i.e., $p_1^2=m_f^2 , p_2^2= m_f^2, q^2 = m_h^2$. We assume that extra Higgs bosons are degenerated in the following calculation. 

In the SM like limit, renormalized scale factor of Yukawa couplings can be approximately expressed as 
 \begin{eqnarray}
 \hat{\kappa}_f = 1- \frac{1}{16\pi^2}\frac{1}{6}\sum_{\Phi=A,H,H^\pm} c_\Phi
 \frac{m_\Phi^2}{v^2}\left(1-\frac{M^2}{m_\Phi^2}\right)^2,
 \label{nonde}
 \end{eqnarray}
where $c_\Phi = 2$ $(1)$ in $\Phi=H^\pm$ $(A, H)$.
The second term in right-hand side of Eq.~(\ref{nonde}) is a deviation from the SM predictions due to loop effects of extra Higgs bosons.
We can see that the effect can be both decoupling and non-decoupling, depending on the balance between $m_\Phi^2$ and $M^2$.
If $M^2$ is as large as $m_\Phi^2$, the effect becomes decoupling in the large mass limit.
Otherwise, quadratic dependences of $m_\Phi$ appear.

\begin{figure}
\begin{center}
\includegraphics[width=70mm]{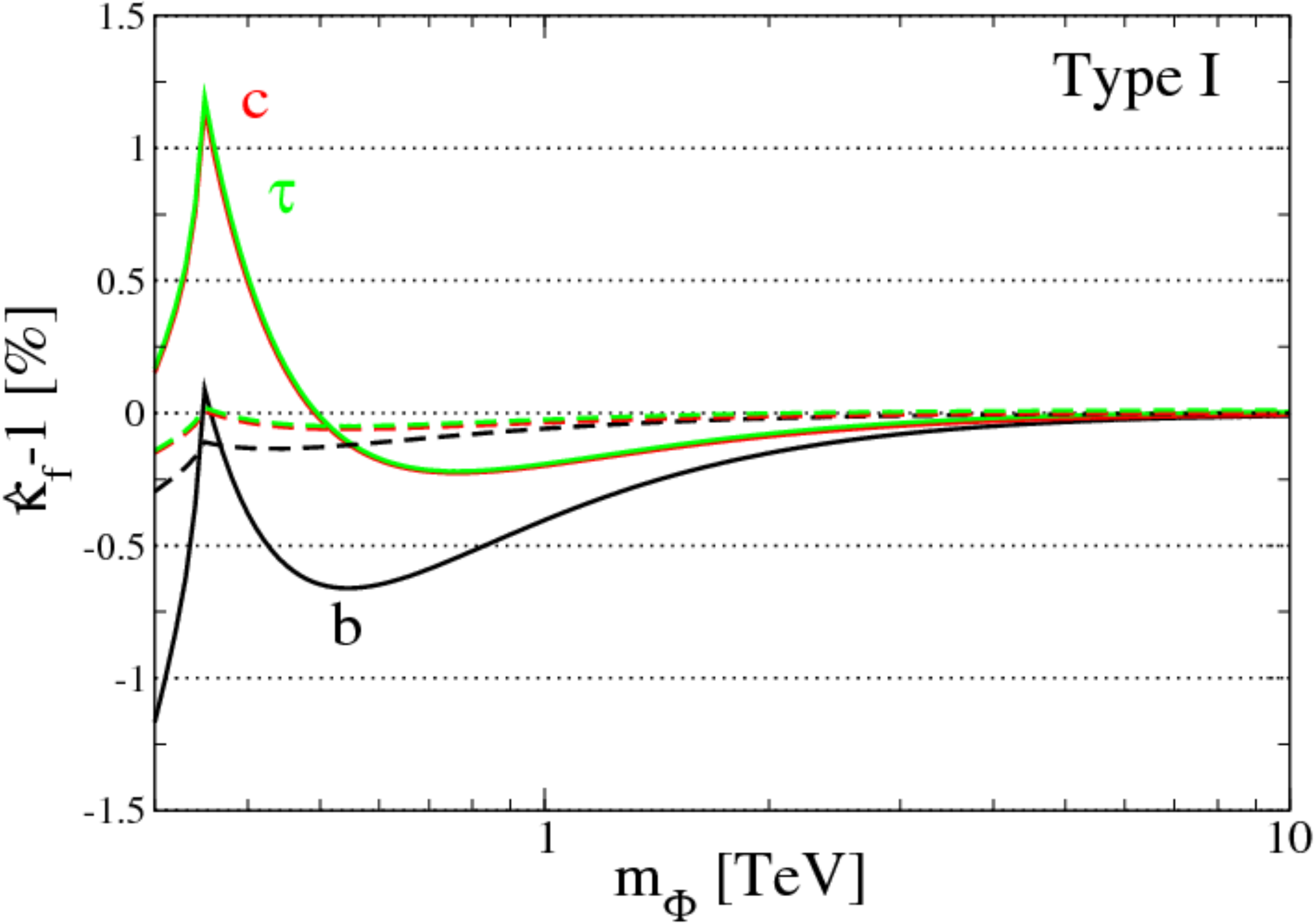}\vspace{1mm}
\includegraphics[width=70mm]{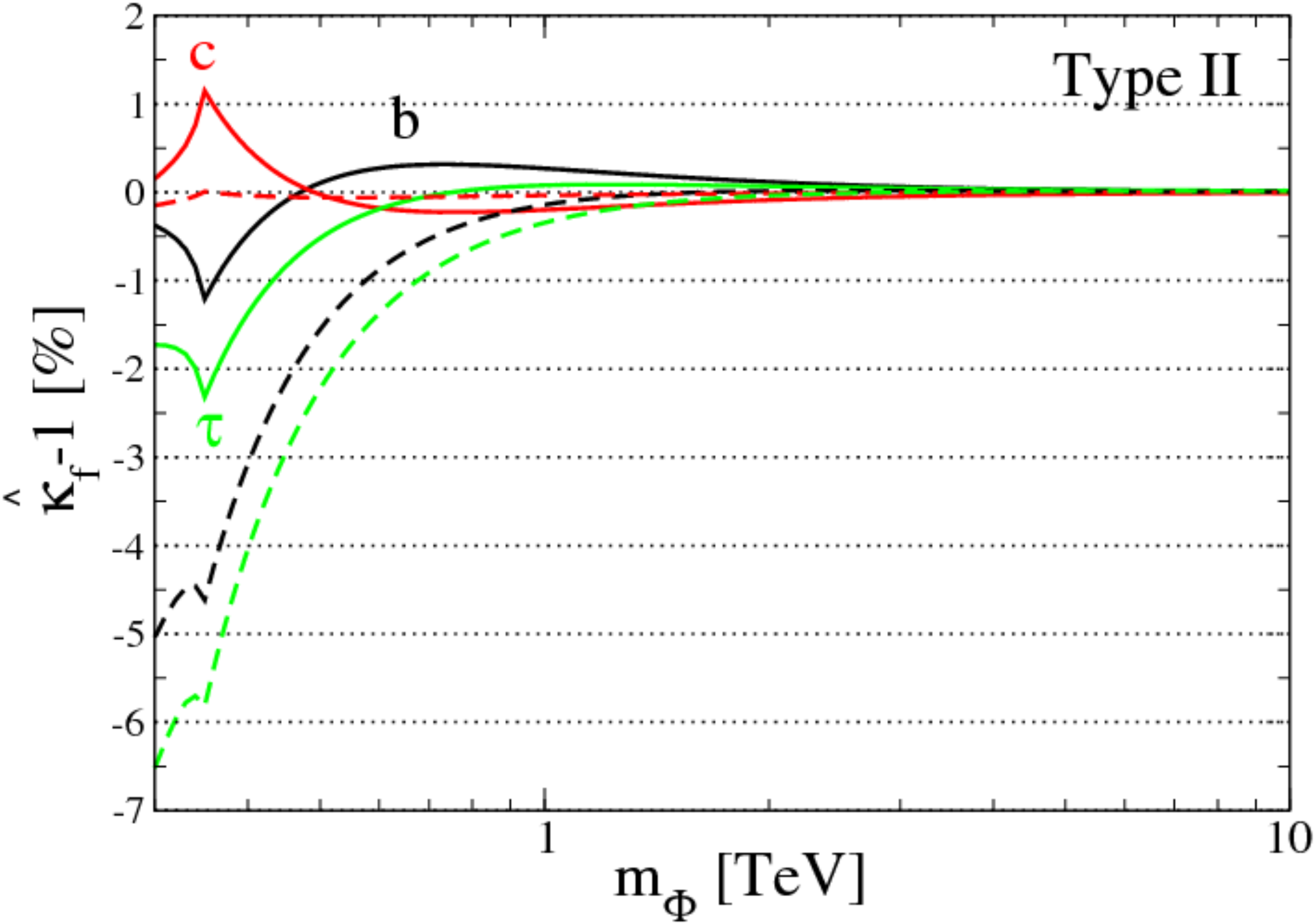}\vspace{1mm} 
\includegraphics[width=70mm]{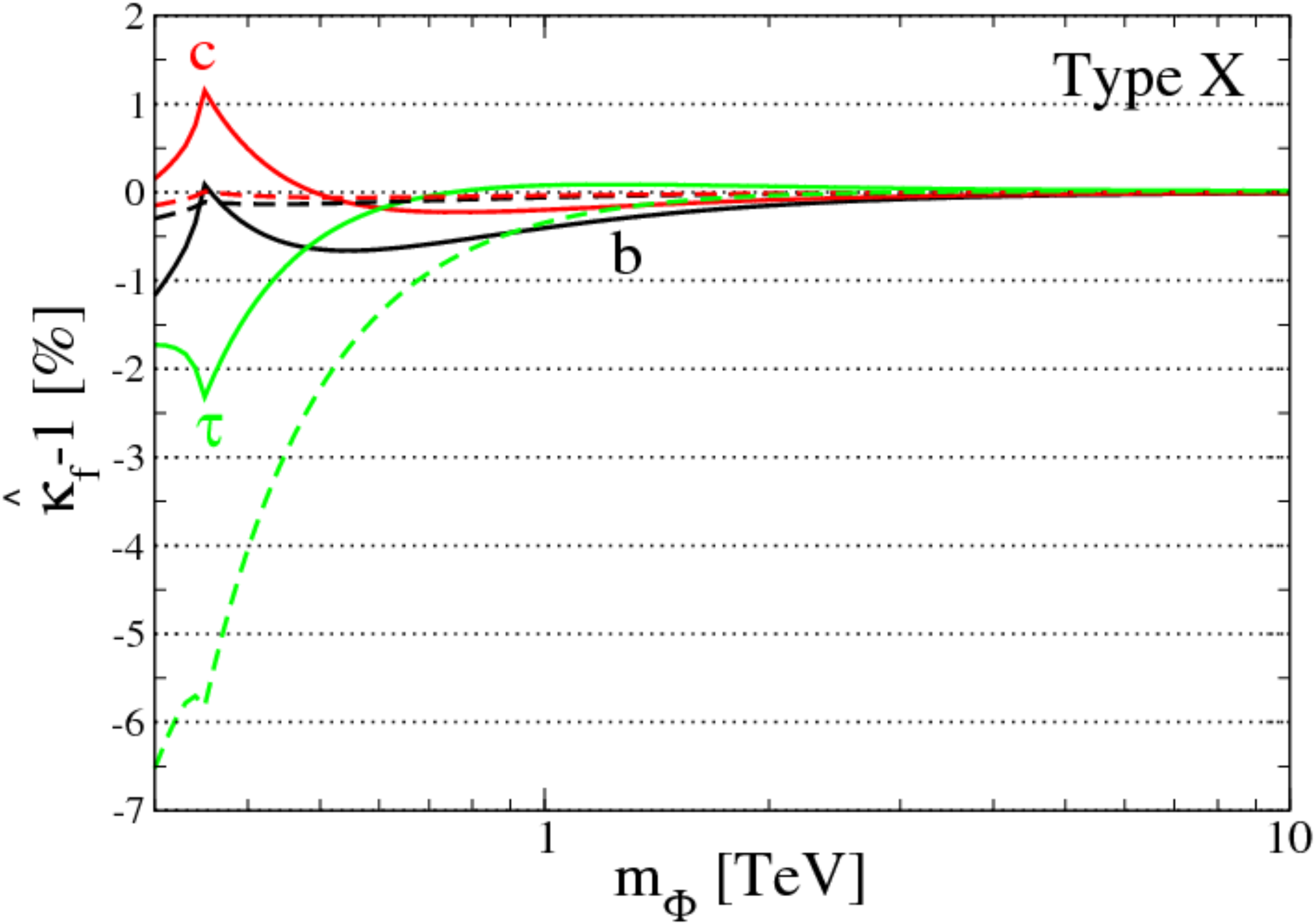}\vspace{1mm}
\includegraphics[width=70mm]{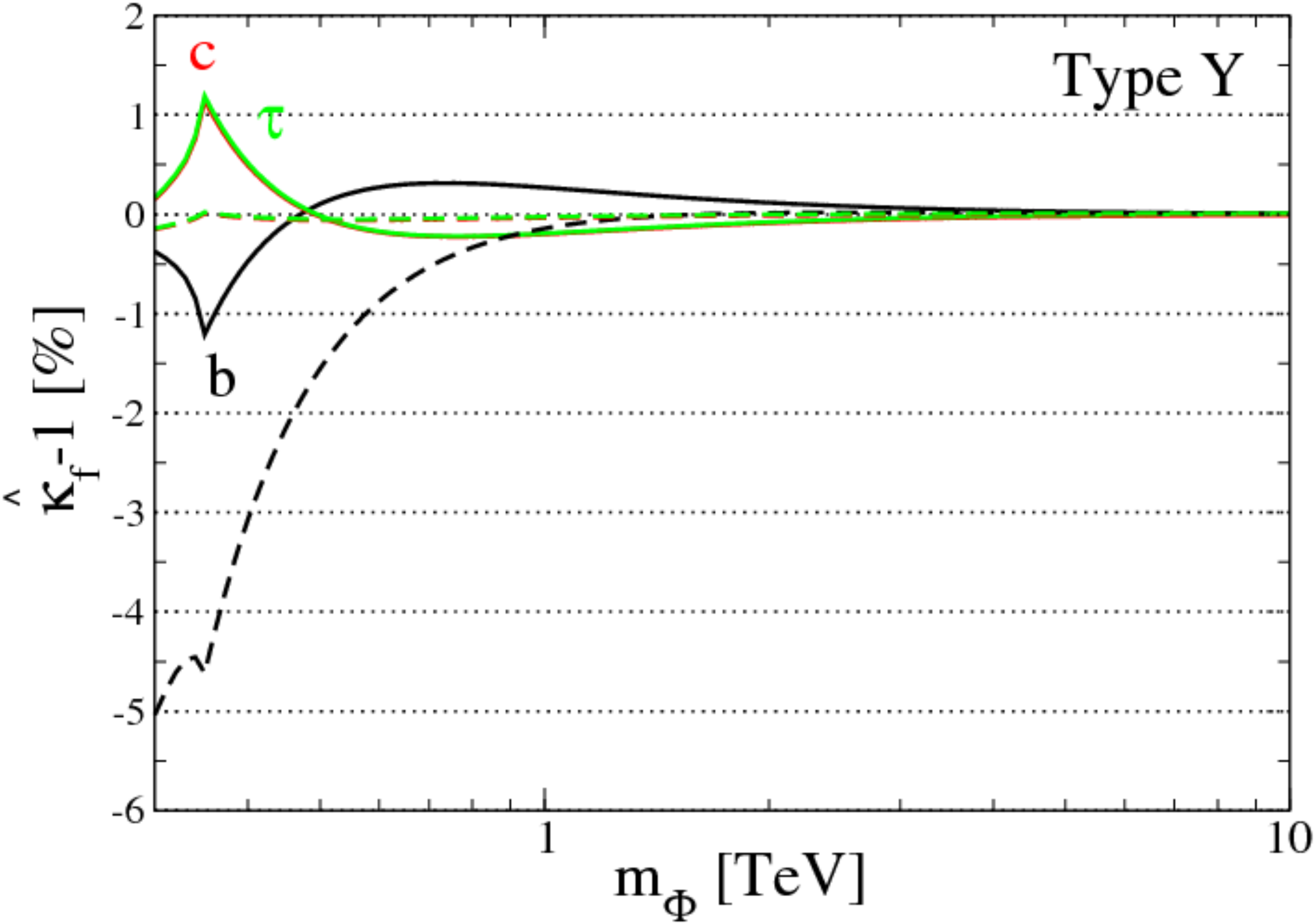}
  \caption{Deviations in $hf\bar{f}$ ($f=b, \tau, c$) couplings in four types of THDMs as a function of $m_{\Phi}$ $(\Phi = H^\pm, A, H)$ when $\sin^2(\beta-\alpha)=1$, $M^2=m_\Phi^2 -(300 \textrm{GeV})^2$~\cite{KKY yukawa}. Solid lines and dashed lines show the case of $\tan\beta=1$ and $\tan\beta=3$, respectively. Those panels show results in Type-I, Type-II, Type-X and Type-Y of THDMs from the top.  }
% Upper-left panel, upper-right panel, lower-left panel and lower-right panel show results in Type-I, Type-II, Type-X and Type-Y of THDMs, respectively.  }
\label{decouple}
\end{center}
\end{figure}
In Fig. 1, we show the decoupling behavior of the one-loop corrections to each Yukawa coupling . 
We plot the deviations in the renormalized Yukawa couplings; i.e., $\hat{\kappa}_f - 1$ for $f = b, \tau$ and $c$ as a function of $m_\Phi$ in the Type-I
(the top), Type-II (the second panel from the top), Type-X (the third one from the top) and Type-Y(the lowest) THDMs with $\sin^2(\beta - \alpha) = 1$, $\tan\beta = 1$ (the solid curves) and $\tan\beta =3$ (the dashed curves). 
We here fix $m_\Phi^2 - M^2$ to be $(300\textrm{GeV})^2$ as just an example.
You notice that the value of the deviations approaches to 0 in the large mass region.
Because $M^2/m_\Phi^2$ gets close to 1 as $m_\Phi$ become larger, the extra Higgs loop contributions written in Eq.~(\ref{nonde}) are reduced.
Thus, we can verify that the renormalized $hf\bar{f}$ couplings approach to the SM prediction in the large mass limit. 
The peak at around $m_\Phi = 2m_t$ is the resonance of the top quark loop contributions to the two point function among $A$ and $G^0$.

\begin{figure}
\begin{center}
\includegraphics[width=70mm]{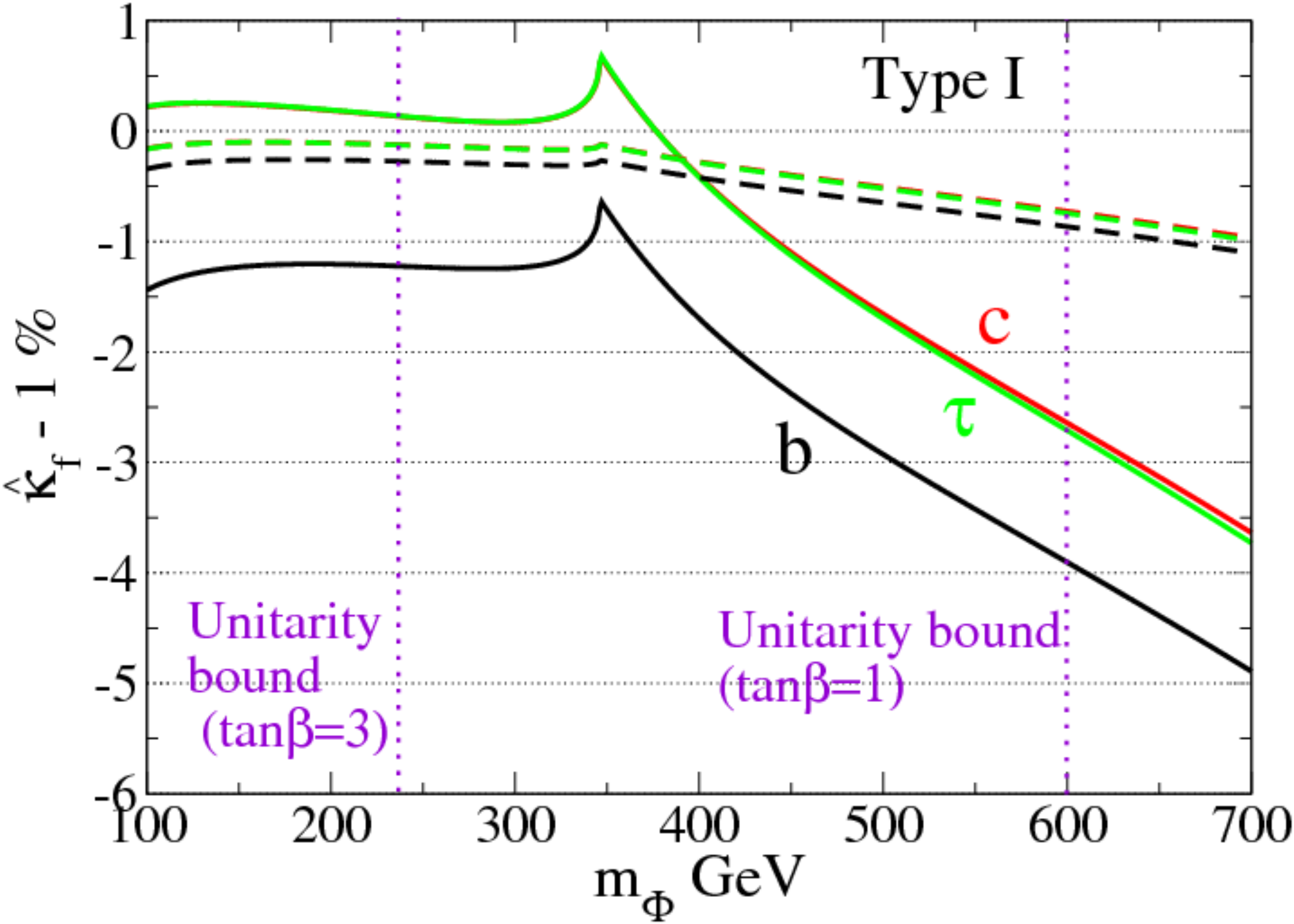}\vspace{1mm}
\includegraphics[width=70mm]{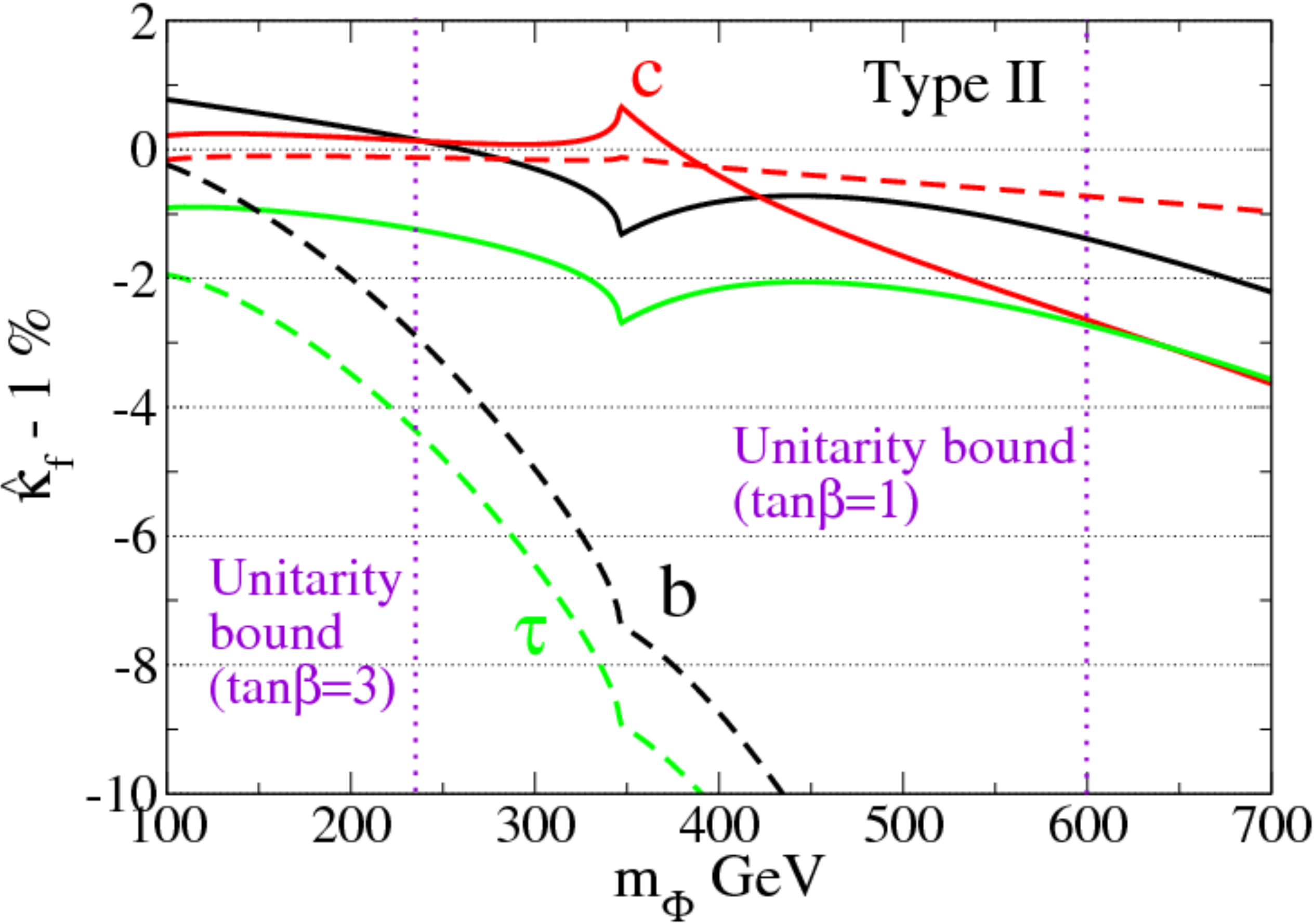}\vspace{1mm} 
\includegraphics[width=70mm]{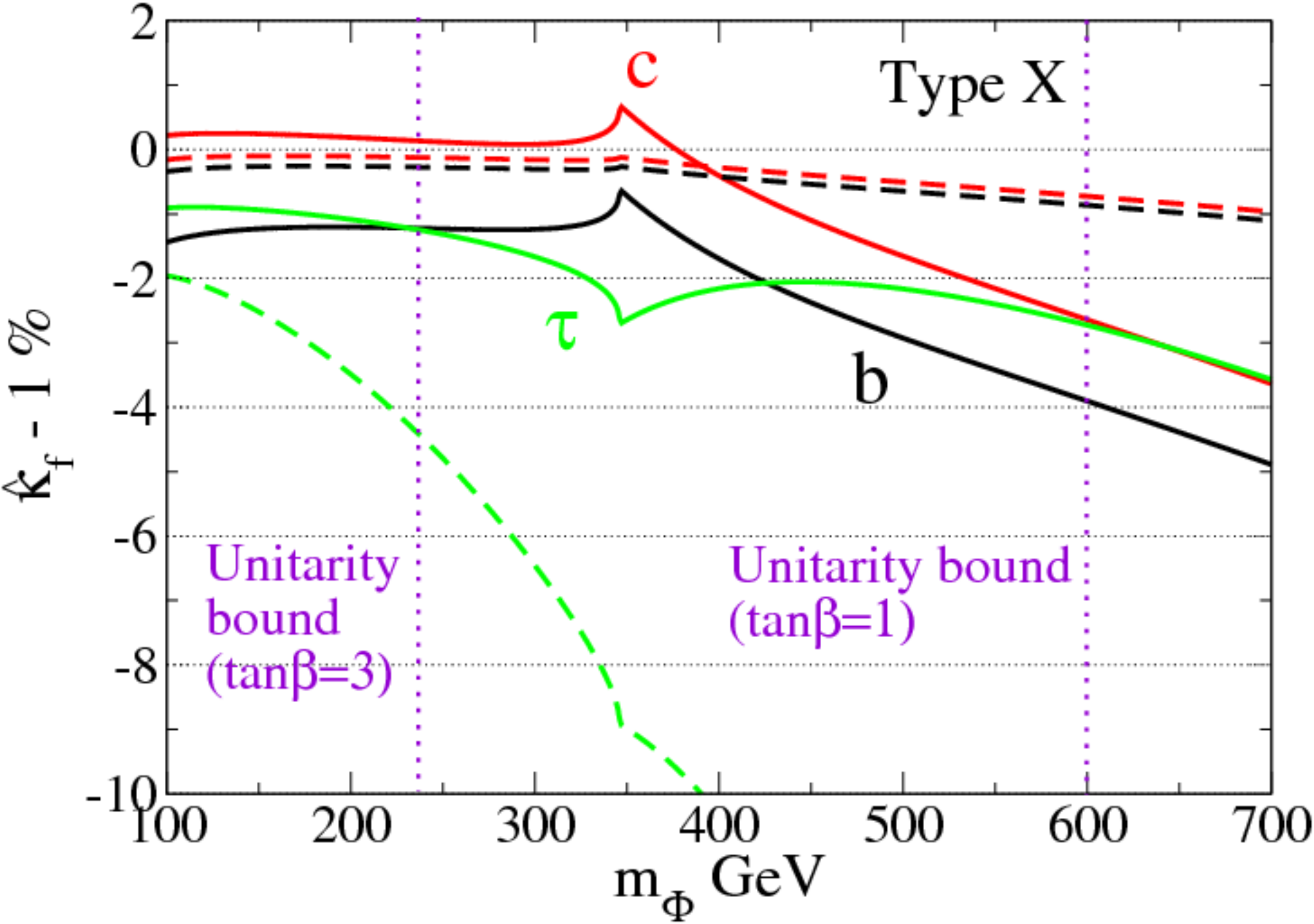}\vspace{1mm}
\includegraphics[width=70mm]{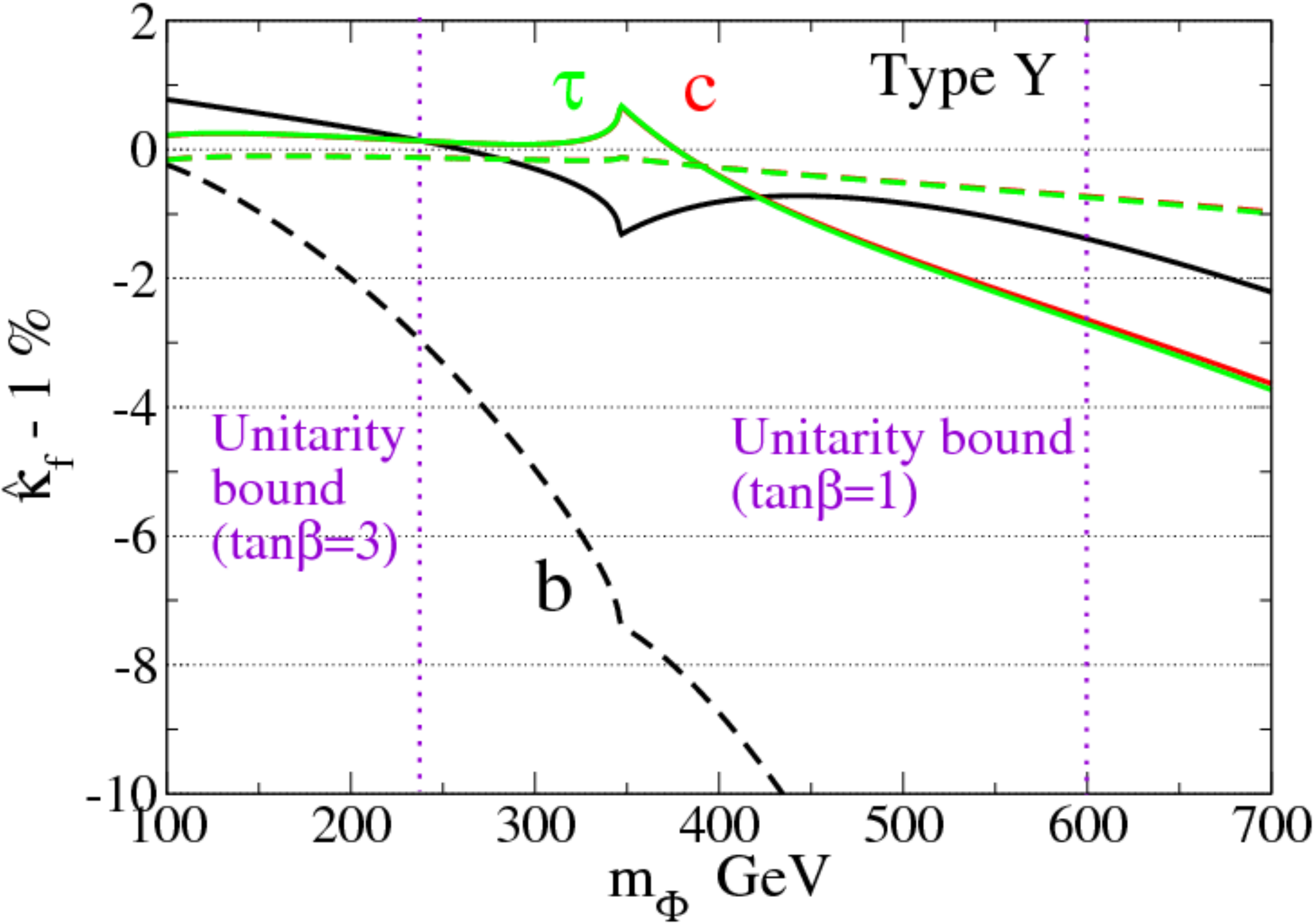}
  \caption{Deviations in Yukawa coupling constants for $b$, $\tau$ and $c$ as a function of $m_{\Phi} $ when $\sin^2(\beta-\alpha)=1$, $M=0$~\cite{KKY yukawa}. Solid lines and dashed lines show the case of $\tan\beta=1$ and $\tan\beta=3$, respectively. 
They are results in Type-I, Type-II, Type-X and Type-Y of THDMs from the top.  }
%Upper-left panel, upper-right panel, lower-left panel and lower-right panel indicate results in Type-I, Type-II, Type-X and Type-Y of THDMs, respectively.  }
\label{decouple}
\end{center}
\end{figure}
In Fig. 2, we discuss non-decoupling effects for deviations in coupling constants of $hcc$, $hbb$ and $h\tau\tau$ in Type-I (the top), Type-II (the second panel from the top), Type-X (the third one from the top) and Type-Y (the lowest).
They are deviations including one-loop radiative corrections as functions of masses of extra Higgs bosons.
We take the mixing angles to be $\sin^2(\beta-\alpha) = 1$ with $\tan\beta = 1$ (solid line) and $\tan\beta = 3$ (dashed line). 
We here fix the value of $M^2$ to be zero. We can find that deviations from the SM predictions can be several percent at the large mass region due to non-decoupling loop effects in the all types of Yukawa interactions even in the case with $\sin^2(\beta-\alpha)=0$. However, the unitarity bound excludes parameter regions where masses of extra Higgs bosons are larger than about $600$ GeV ($230$ GeV) in $\tan\beta=1 (3)$.

%% \begin{figure}
%% \begin{center}
%% \includegraphics[width=80mm]{nonde300II}
%%   \caption{Deviations of Yukawa coupling constants in Type-II of 2HDMs as a function of $m_{\Phi} $when $\sin^2(\beta-\alpha)=1$, $M=3000\textrm{GeV}$. Solid lines and dash lines show the case of $\tan(\beta)=1$ and $\tan(\beta)=3$, respectebility.} 
%% \label{decouple}
%% \end{center}
%% \end{figure}
%% In FIG. 2, we numerically calculate $\hat{\kappa}_f - 1$ in Type-II taking $M = 300$ GeV.
%% Parameter sets are same as those in Fig. 1 except $M$.
%% In this case, we find that the degree of non-decoupling become slightly smaller than the one in the case with $M=0$.  

\begin{figure}
\begin{center}
\includegraphics[width=110mm]{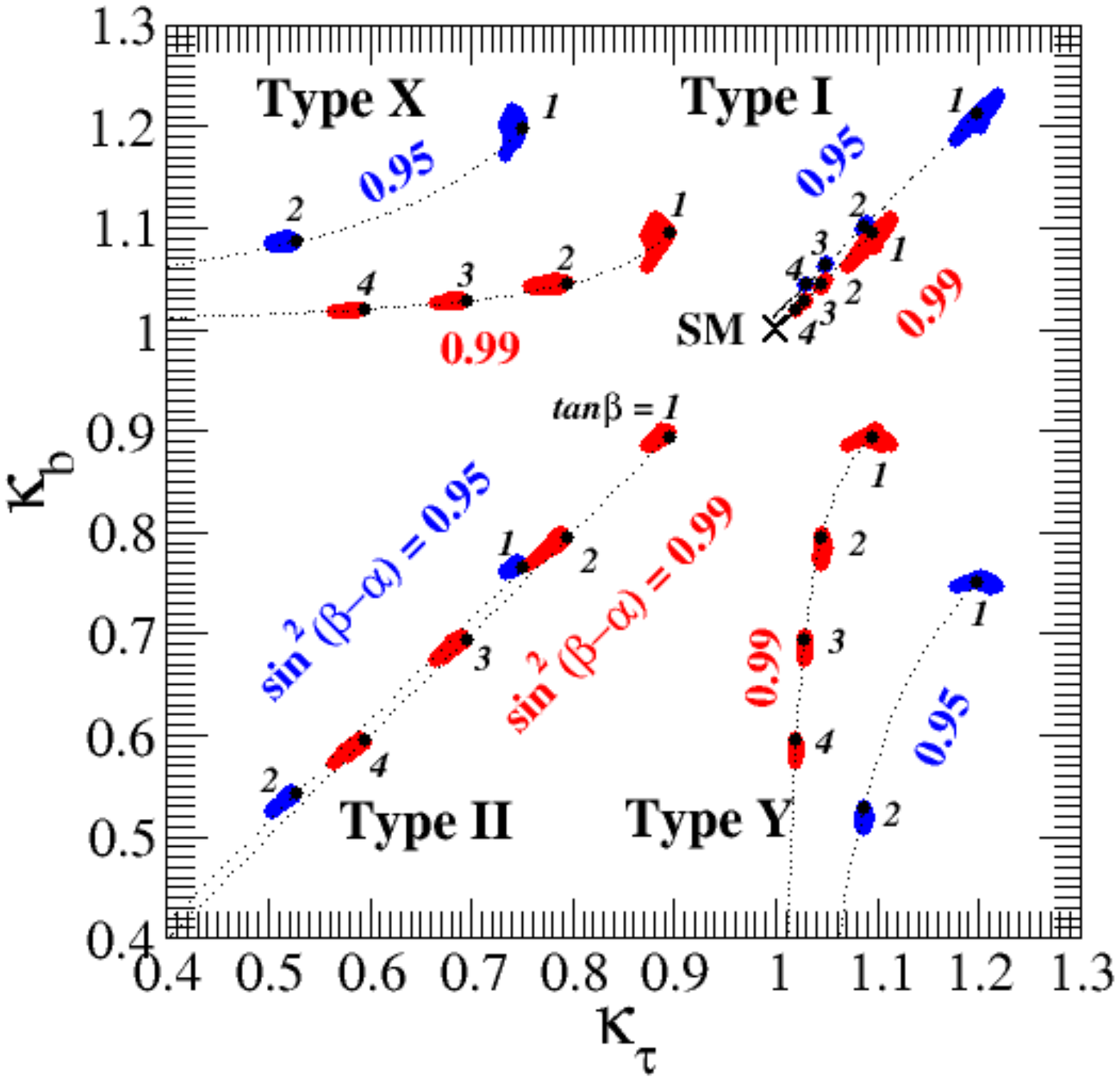}\vspace{1mm} 
\includegraphics[width=110mm]{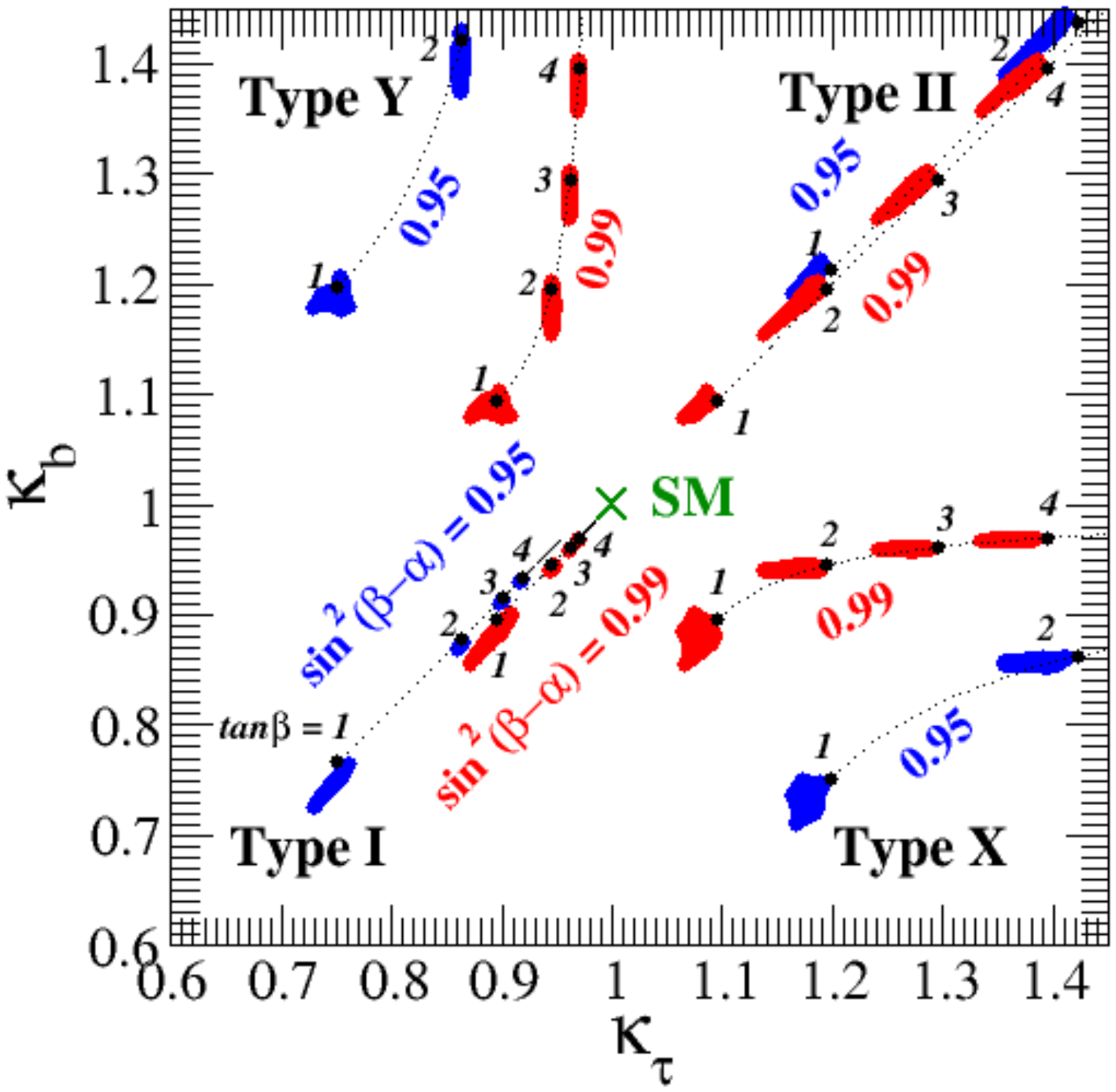}
\caption{Plots of scale factors of $\tau$ and $b$ in four types of Yukawa interactions~\cite{KKY yukawa}. 
The upper panel and the lower panel are predictions with $\cos(\beta-\alpha)<0$ and $\cos(\beta-\alpha)>0$, respectively. 
Each black dot indicates a result at the tree level with $\tan\beta = 1,2,3$ and $4$. 
Red region (blue region) show one-loop results with $\sin^2(\beta-\alpha)=0.99$ ($\sin^2(\beta-\alpha)=0.95$) where $m_\Phi$ and $M$ are scanned over from 100 GeV to 1 TeV and 0 to $m_\Phi$, respectively, under the constraints of perturbative unitarity and vacuum stability.} 
\label{finger1}
\end{center}
\end{figure}

In Fig. 3, we show the behavior of the scale factors at the tree level $\kappa^\textrm{tree}_\tau$, $\kappa^\textrm{tree}_b$ and one-loop corrected scale factors $\hat{\kappa}_\tau$, $\hat{\kappa}_b$ in the four types of THDMs~\cite{KKY yukawa}. 
The upper panel and the lower panel correspond to results in the case with $\cos(\beta-\alpha) < 0$ and $\cos(\beta - \alpha) > 0$, respectively. 
Doted lines indicate predictions at the tree level in $\sin^2(\beta-\alpha) = 0.99$ and $0.95$, and black dots being on these lines are the tree level results with $\tan\beta = 1, 2, 3$ and $4$. 
At the tree level, in the case with $\sin^2(\beta-\alpha) = 1$, predictions of all the types get close to those of the SM. 
If $\sin(\beta-\alpha)$ slightly deviate from unity, $\kappa^\textrm{tree}_f$ for each type lead to deviate in different directions.
However it is diffecult to discriminate the types of THDMs by evaluating only $\kappa_b$ and $\kappa_\tau$ because behaviors of $\kappa_b$ and $\kappa_\tau$ depend on the sign of $\cos(\beta-\alpha)$.
If $\cos(\beta-\alpha)$ is negative (positive), predictions of $\kappa_{t(c)}$ in all the types are less (larger) than 1.
Therefore we can determine the sign of $\cos(\beta-\alpha)$ by using measurements of $\kappa_{t(c)}$.  
Then we can discriminate all types of Yukawa interactions by the pattern of deviations in these $hf\bar{f}$ couplings.
These analysis of Yukawa couplings at the tree level have already been discussed in Refs. \cite{ILC white, fingerp, kanemura}.

In Fig.~\ref{finger1}, we also plot those including full electroweak and scalar bosons loop corrections which are shown by colored regions around black dots. 
Red regions (blue regions) are modified regions by extra Higgs loop contributions for the case with $\sin^2(\beta-\alpha) = 0.99$ ($0.95$). 
We scan $m_\Phi(=m_{H^\pm}=m_A=m_H)$ and $M$ over from 100 GeV to 1 TeV and from $0$ to $m_\Phi$, respectively. 
We find that results can be modified from the tree level values
in several percent by extra Higgs loop effects.
%\textcolor{red}{ 
%In the case with $M= $, radiative corrections become maximal by non-decoupling effects due to extra Higgs bosons loop. }
Even if radiative corrections become maximal values, predictions of $\hat{\kappa}_f$ ($f = c, b, \tau$) in the types of Yukawa interaction don't overlap each other.
Therefore we can discriminate all the types 
when $\sin^2(\beta -\alpha)$ deviates from the SM prediction by about 1$\%$.

At the HL-LHC, $h\tau\tau$ and $hbb$ couplings are expected to be measured with about $8\%$ and $11\%$, respectively ~\cite{CMS note}. 
When $\sin^2(\beta-\alpha)$ is different about $1\%$ from unity, 
$hbb$ and $h\tau\tau$ coupling constants can differ about $10\%$ from the predictions of the SM depending on the value of $\tan\beta$.  
In that case, we can discriminate the types of Yukawa interactions by using those HL-LHC data. 
At the ILC500, however, the Higgs coupling measurements have typically $\mathcal{O}(1)\%$ level resolution: e.g., $h$ coupling constants to $\tau$ and $b$ can be determine with $2.3\%$ and $1.6\%$ uncertainty, respectively 
%in the version with $\sqrt{s}=500$ GeV and $L=500 \textrm{fb}^{-1}$ 
~\cite{ILC white}. 
In order to compare with such precision coupling measurements at the ILC, 
we must not neglect the effects of radiative corrections.

 \section{Conclusion}
In extended Higgs models, properties of each model appear as the pattern of deviations in SM-like Higgs boson couplings from those in the SM. 
In four types of THDMs with the softly-broken $Z_2$ symmetry, $hf\bar{f}$ couplings deviate from the predictions in the SM by different patterns each other.
Therefore there is the possibility to discriminate all the types by those correlate relations among $hf\bar{f}$ couplings. 
On the other hand, it is expected that $h$ coupling constants are measured typically by $\mathcal{O}(1)\%$ at the ILC. 
In order to compare theoretical predictions with such high precision data, we evaluate Higgs couplings with radiative corrections. 
We calculate a full set of loop corrections for electroweak sector and the scalar sector by the on-shell renormalization scheme.
We have found that each Yukawa coupling can modify about several percent from the tree level prediction by extra Higgs loop corrections. 
These differences are not negligible to compare the ILC precision measurements. If gauge couplings, such as $hWW$ and $hZZ$, slightly deviates from the SM predictions enough to measure at the ILC, predictions of $\hat{\kappa}_\tau$ and $\hat{\kappa}_b$ in all the types do not overlap each other even in the case with maximal radiative corrections and we can distinguish the type of the THDM.

\section*{Acknowledgments}
%I'm very grateful to Shinya Kanemura and Kei Yagyu for the wonderful collaborations of this work. 
This work was supported by Grants-in-aid for JSPS, No. 25$\cdot$10031.



%% References

%\nocite{*}
\bibliographystyle{elsarticle-num}
\bibliography{martin}
%\noindent

%\end{acknowledgments}



\end{document}